# The Casimir effect at the nucleus

By Frank Kowol, Nov. 2023

1. **Abstract**

In this report, the impact of the Casimir effect in the near-nuclear environment on electrons, in particular of the K-shell, is investigated. It has long been known that the experimentally measured binding energies of the inner electrons, especially for heavy elements, agree only to a very limited extent with the theoretical solutions of the Schrödinger and Dirac equations. This report presents a modification for the potential of electrons near the nucleus by investigating the impact of the Casimir effect on the innermost electrons. It can be shown that with this approach the calculated binding energies agree much better with the literature values from spectroscopy, especially for heavy elements.

In addition, the innermost electrons are apparently influenced much more strongly than previously assumed by the nature of the nuclear surface, in particular the deviation from the spherical geometry, e.g. the multipole moments and not least the range of the strong interaction. This effect is not to be confused with the quantum-mechanical quadrupole energy (Casimir 1936), which is determined by spin interactions between nucleus and electron, but the effect discussed in this report is a distance-dominated effect of the s-electrons to the nucleus. It offers the possibility of investigating the nucleus structure and isotope effects on the binding energies much more precisely than expected, e.g. by spectroscopy, and thus gaining new insights into the respective nuclear structure and geometry.

At the same time, this approach shows that the probability of the electrons staying close to the nucleus is apparently significantly higher than assumed in previous atomic models. This may also increase the transition probabilities of electron capture, and thus the model enables a higher accuracy for the calculation of the theoretical half lifetimes for electron capture decay.

## 2. Derivation of the Casimir-effect

Originally regarded as an obscure effect of quantum electrodynamics, the Casimir effect is now theoretically very well understood and has been proven experimentally many times over [1,3,4,5,6,7,8,9,10,11,12]. It describes the relativistic van der Waals forces initiated by vacuum fluctuations that can occur between conductive surfaces [3]. The original - and theoretically easiest to motivate - experimental setup describes the attractive distance-dependent interaction effect of two plane-parallel conductive surfaces. The basic prerequisites for this are polarizable charge distributions bound to a stationary region and the existence of a defined surface. Thus, the Casimir effect can not only be investigated for conductive surfaces, e.g. in micromechanics, but can also be extended to stationary electric charge distributions, such as stationary plasmas or bound electrons in atoms, provided that a surface can be assigned to each of these charge distributions [4]. Consequently, this also applies to the charge distributions of the protons in the nucleus. To stay with the image of the original Casimir effect: One can consider the nucleus shell and the differential contributions of the radial wave function of the 1s electron as the two surfaces of a spherical capacitor and thus calculate the effects on the electrons. With regard to the originally required freedom from charge in the original set-up, the following can be said in reply: In order to obtain a precise measurement result from this very delicate experiment, it is of course necessary to ensure that there is no parasitic force, such as a Coulomb interaction, as this would mask the Casimir effect. For the basic mechanism of induced polarization, however, the requirement for freedom from charge is neither necessary nor sensible.

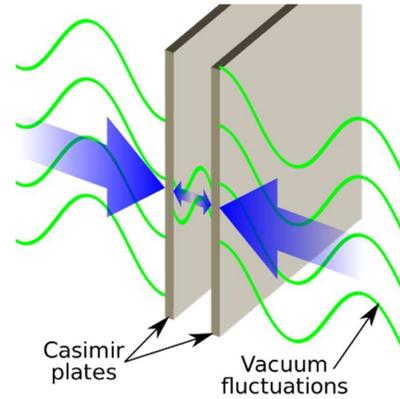

Fig.1: Casimir effect between two plane-parallel plates. From [53]

In order to capture the Casimir effect mathematically, as it was originally derived by Casimir and Polder [3], a summation over the states of the quantized electromagnetic field in the form of quantum mechanical harmonic oscillators is carried out as known from quantum electrodynamics. The following then applies to the energy contribution and the resulting force:

$$\langle E \rangle = \frac{1}{2} \sum_n E_n \quad \text{mit} \quad F(x) = -\frac{\delta \langle E(S) \rangle}{\delta S}\bigg|_x \qquad (1)$$

Where the n modes of energy $E = \frac{1}{2}\hbar\omega$ are summed up here. $\langle E(S) \rangle$ corresponds here to the expected value of the zero-point energy, depending on the shape of the boundary surface S and F(x) is the resulting force.

For the simplified case of plane-parallel plates, especially if the distance between the plates is significantly smaller than the plate dimensions, the electromagnetic field of the virtual particles can be represented as follows (with the plates parallel to the x-y plane):

$$\psi_n(x,y,z,t) = e^{-i\omega_n t} e^{i(k_x x + k_y y)} sin(k_n z)$$

The wave vectors $k_x$ and $k_y$ are plane-parallel to the plates The wave vector $k_n = n\pi a^{-1}, n \in \mathbb{N}$ orthogonal to the plates results in a wave whose amplitude vanishes at the points where the plates touch. The distance a indicates the distance between the two plates, and the frequency of the wave corresponds to (with c as the speed of light):

$$\omega_n = c\sqrt{k_x{}^2 + k_y{}^2 + \left(\frac{n\pi}{a}\right)^2} \qquad (2)$$

To determine the resulting energy, it is necessary to integrate over two dimensions in k-space and both possible polarizations of the wave. If periodic boundary conditions are assumed, then with A as the area for the vacuum energy and Planck's reduced constant $\hbar$:

$$\frac{\langle E(s)\rangle}{A} = \hbar \int \frac{dk_x dk_y}{4\pi^2} \sum_{n=1}^{\infty} \omega_n |\omega_n|^{-s} \qquad (3)$$

The expression $|\omega_n|^{-s}, s \in \mathbb{C}$ is a regulator with $\lim_{s \to 0} |\omega_n|^{-s} = 1$, as the expression otherwise tends to infinity. After some complex calculation using the Riemann zeta function, it can be shown, that

$$\frac{\langle E\rangle}{A} = \lim_{s \to 0} \frac{\langle E(s)\rangle}{A} = -\frac{\hbar c \pi^2}{6a^3}\zeta(-3) \ \ mit \ \zeta(-3) = \frac{1}{120} \qquad (4)$$

The following therefore applies to the energy per area and the Casimir pressure (= force per area) [3,4]

$$\frac{\langle E\rangle}{A} = -\frac{\hbar c \pi^2}{720 a^3} \qquad (5)$$

and

$$\frac{\langle F\rangle}{A} = -\frac{d}{da}\frac{\langle E\rangle}{A} = -\frac{\hbar c \pi^2}{240 a^4} \qquad (6)$$

Equation 6 already shows that the range of the Casimir effect only provides significant contributions at very short distances, that is it only becomes perceptible in the immediate vicinity of the interface. In micromechanics and with intermolecular effects, (6) provides a measurable contribution and has been thoroughly investigated [5 - 12].

### 3. The Casimir effect for electrons close to the nucleus

To avoid confusion, we are not talking here about the quadrupole energy as derived by Casimir in 1936. Referring to the derivation according to [2], this is as follows:

$$E_Q = \frac{1}{4}\left(\frac{\partial E_z}{\partial z}\right)_0 eQ\frac{\frac{3}{2}C(C+1) - 2I(I+1)J(J+1)}{I(2I-1)J(2J-1)} \ with \ F(F+1) \ - \ I(I+1) \ - \ J(J+1)$$

With F, I, J as quantum numbers (F for the magnetic splitting, I as nuclear spin and J = L+S for the electron spin) As can easily be seen, $E_Q$ is singular with $J = 0, \frac{1}{2}$, so electrons of the K-shell singlet cannot be considered by the above formula.

However, if we look at the atomic region near the nucleus, we can use $a = r - r_k$ with radius $r > 0$ and $r_k$ as the nuclear radius to investigate the well-known Casimir effect between the electron wave function and the nuclear surface. Now two questions immediately arise: How to deal with the singularity at $r \to r_k$, and how to interpret areas inside the nucleus with $r < r_k$.

It is important to note that the nucleus in these orders of magnitude of <10 femtometers has no longer an ideal two-dimensional boundary surface. The original approach from the derivation of the Casimir energy was based on plane-parallel plates with the density function $\varrho(r) = \delta(r - r_{shell})$. In fact, however, this is unphysical in this case, because we are dealing with protons in a finite potential well, which according to quantum mechanics have an exponentially decreasing probability of residence for $r > r_k$. Within the potential well, they would be delocalized, that means the interactions and laws of quantum chromodynamics are applied. However, the electron only perceives (exchangeable) protons, or rather their shares via the boundary surface. In the nucleus, the electron would in turn be surrounded by arbitrarily interchangeable protons, provided the spin-spin interactions are disregarded.

The following conclusion can therefore be drawn: A legitimate approach is to contract the charge of the nucleus on the edge of the nucleus with spherical symmetry and thus define the interior of the nucleus as field-free - at least for the electron. The diffuse nature of the boundary can then be considered by describing the probability of the proton's position with

$$f(r) := |\psi_P(r)|^2 = \lambda e^{-\frac{(r-r_k)^2}{2(\varepsilon\alpha)^2}} \ \ with \ \ \alpha = \frac{2\hbar}{\sqrt{2m_P E_B}} \ \ and \ \varepsilon > 0 \ \ as \ well \ as \ \int_{\mathbb{R}} f(r)dr = 1 \qquad (7)$$

$m_P$ corresponds to the rest mass of the proton and $E_B$ to its binding energy for the respective isotope, typically in the order of 7-9 MeV. $\lambda = const$ is the normalization factor. Of course, $\lim_{r \to 0} \psi_P(r) = 0$ and $\lim_{r \to \infty} \psi_P(r) = 0$ also apply. The parameter $\varepsilon$ can be interpreted as a convergence-generating factor, because in the transition for small $\varepsilon$ $\psi_P$ merges into the delta function: $\lim_{\varepsilon \to 0} \psi_P(r) \to \delta(r - r_k)$.

If one follows the idea of an edge blurring of the nucleus, then $\varepsilon$ has another meaning, namely the parameterized blurring of the nucleus surface. With the definitions [13]:

$$\langle r \rangle = \frac{1}{4\pi} \int\limits_0^{2\pi} d\varphi \int\limits_0^{\pi} r(\theta, \varphi) \, sin\theta \, d\theta \;\; and \;\; \sigma_r^2 = \int\limits_0^{2\pi} d\varphi \int\limits_0^{\pi} (r(\theta, \varphi) - \langle r \rangle)^2 \, sin\theta \, d\theta \tag{8}$$

$r(\theta, \varphi) = r_k \left(1 + \delta r_{l,m} Y_{l,m}(\theta, \varphi)\right)$ can be understood as a function for the deviation of the nucleus from spherical symmetry [46]. In the case of an ideal sphere, $r(\theta, \varphi) = r_k = const.$ and hence $\langle r \rangle = r_k$ and $\sigma_r^2 = 0$. This results in the following relationship:

$$\varepsilon = 1 + \frac{1}{\alpha} \sum_{l=0}^{\infty} \sum_{m=-l}^{l} \delta r_{l,m} \left( \int\limits_0^{2\pi} d\varphi \int\limits_0^{\pi} \left( Y_{l,m}(\theta, \varphi) \right)^2 sin\theta \, d\theta \right)^{\frac{1}{2}} = 1 + \frac{1}{\alpha} \sum_{l=0}^{\infty} \sum_{m=-l}^{l} \delta r_{l,m} \tag{9}$$

The occurring multipole moments can now be considered in the distribution function. However, it should be noted that the monopole moment is already contained in the nuclear charge Z and the nucleus shows no dipole moment for reasons of parity conservation [31]. The first significant contribution is the electric nuclear quadrupole moment $Q_b > 0$ (given in barn, 1barn = $10^{-28}$m$^2$). The image of a prolate (cigar-shaped, $Q_b > 0$) or oblate (flattened, $Q_b < 0$) rotational ellipsoid is often used. In the literature, this is identified with a deformation parameter; taking equation (9) into account, this can therefore be written as [31, 43]:

$$\varepsilon = 1 + \frac{1}{\alpha} \frac{5}{4} \frac{Q_b}{Ze} \frac{1}{\langle r_k \rangle} \tag{10}$$

If we now combine equation (7) with the findings from equation (10) - with the quadrupole moment as a first approximation - the question arises as to how the distributed structure of the interface can be captured mathematically. One way to do this would be to use the convolution theorem [13,14,25,28]:

$$\{f * F\}(r) = \int\limits_0^{\infty} f(r')F(r - r')dr' = -\lambda \frac{\hbar c \pi^3}{60} r_k^2 \int\limits_0^{\infty} \frac{e^{-\frac{(r\prime - r_k)^2}{2(\varepsilon\alpha)^2}}}{(r - r')^4} dr' \tag{11}$$

The Casimir effect only provides contributions for radial segments of the electron $|\psi_e(r)|^2 r^2 dr$, which are very close to the nuclear surface, so that the original spherical symmetry of the interaction zone can be converted into a linearized form of plane-parallel plates with the distance $r - r_k$. Accordingly, the area A, to which the Casimir pressure refers, is identified with the mean core surface: $A = 4\pi r_k{}^2$. This integral can be solved using the convolution theorem and is described in detail in the Annex. The result is:

$$F_{Casimir}(r) = -\frac{\hbar c \pi^3 r_k{}^2}{240(\sqrt{2}\varepsilon\alpha)^7} e^{-\frac{(r-r_k)^2}{2(\varepsilon\alpha)^2}} |3(\varepsilon\alpha)^2(r-r_k) - (r-r_k)^3|$$

(12)

The analytical representation of $V_{Casimir}(r)$ is defined piecewise due to the occurring sgn function, as listed in the annex. The result can be seen in Figure 2. However, the substructures on the edge of the potential $V_{Casimir}(r)$ can lead to numerical artifacts when solving the Dirac equation, which is why the almost identical error function $Erf(r)$ was used as a first approximation as the primitive of the "Gaussian bell". For this purpose, the error function was calibrated accordingly regarding slope and potential depth as shown in the figure to calculate the wave function. However, for the energy analysis the original Casimir-potential was used. The numerical analysis with the potential $V'_{Casimir}$ obtained in this way shows stable and evaluable results for all elements. The effective Casimir potential for the electrons close to the nucleus is shown in equation (13).

A closer look at the potential reveals the effect of the Casimir effect on the electron; the additional potential term obviously has a broadening effect on the original Coulomb field. At $r \gg r_k$, the electron will experience almost no effect, so the lighter elements with low atomic numbers and the electrons further away from the nucleus will remain unaffected by this mechanism.

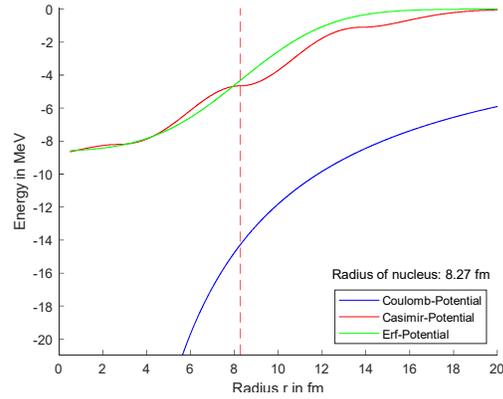

Fig.2: Coulomb-, Casimir- and calibrated Erf potential for $Pb_{82}^{206}$ in MeV plotted against the nuclear distance in fm. The nucleus radius was assumed to be 8.27 fm. Source: Calculation by Matlab-Script

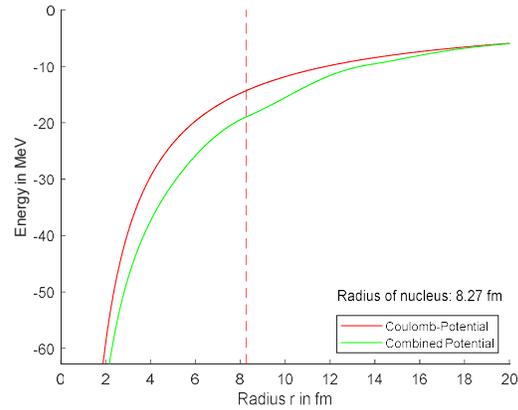

Fig.3: Pure Coulomb potential and combined potential with Coulomb and Casimir part $V_{Casimir}$ for $Pb_{82}^{206}$ in MeV plotted against the nuclear distance in fm. The nucleus radius was assumed to be 8.27 fm Source: Calculation by Matlab script

For higher Z, the electrons of the K shell will be much closer to the nucleus due to the Coulomb force

$$V'_{Casimir}(r) = -\frac{\hbar c \pi^3 r_k{}^2}{240(\sqrt{2}\varepsilon\alpha)^3}\left(e^{-\frac{3}{2}} + \frac{1}{4}\right)\left[Erf\left(\frac{r-r_k}{\sqrt{2}\varepsilon\alpha}\right) - 1\right]$$

(13)

alone. The Casimir effect will bind the electrons even closer to the nucleus as an additional force term.

Finally, for $r < r_k$, $V'_{Casimir}$ as well as $V_{Casimir}$ changes into a box potential and will have a corresponding influence on the binding energy of the electrons. To evaluate this precisely, the Dirac equation must be solved.

### 4. The Dirac-equation for the Coulomb-potential

In contrast to the Schrödinger equation, the Dirac equation already includes relativistic effects and the electron spin, which is why it is typically formulated with the help of spinors or alternatively written in covariant form [15,16,17,18,19,20,21,22,23,24,26]. In contrast to the Schrödinger equation, the eigenvalues of the energies already contain the rest energy term $\pm mc^2$ for both particles and antiparticles, as both particles represent a valid solution of the Dirac equation. The derivation and detailed consideration of the equation can be found elsewhere in much more detail and in a better didactic form [24,26], which is why only the solution for particles in the Coulomb potential is outlined here. The general form of the Dirac equation reads as

$$i\frac{\partial \psi}{\partial t} = [c\boldsymbol{\alpha} \cdot \hat{\boldsymbol{p}} + \boldsymbol{\beta} m_0 c^2 + V]\psi = H_D \psi \qquad (14)$$

where $\boldsymbol{\alpha}$ and $\boldsymbol{\beta}$ are 4 x 4 matrices and $\hat{\boldsymbol{p}} = -i\nabla$ is the impulse operator. Hence $\boldsymbol{\alpha}$ can be written with the Pauli-Matrices

$$\sigma_x = \begin{pmatrix} 0 & 1 \\ 1 & 0 \end{pmatrix}, \sigma_y = \begin{pmatrix} 0 & -i \\ i & 0 \end{pmatrix}, \sigma_z = \begin{pmatrix} 1 & 0 \\ 0 & -1 \end{pmatrix}$$

with

$$\alpha_{x,y,z} = \begin{pmatrix} 0 & \sigma_{x,y,z} \\ \sigma_{x,y,z} & 0 \end{pmatrix}$$

While $\boldsymbol{\beta}$ can be expressed with the 2 x 2 identity matrix $\mathbb{1}$

$$\beta = \begin{pmatrix} \mathbb{1} & 0 \\ 0 & -\mathbb{1} \end{pmatrix}$$

The solution in the central potential with $V(r) = -\frac{Ze^2}{4\pi\varepsilon_0}\frac{1}{|r|}$ is carried out by separating the variables using similar approaches as for the Schrödinger equation, but in the Dirac equation one ends up with two time-independent coupled differential equations (for particles and antiparticles), where each contains the first derivatives with respect to the location. The conservation variables here are: J = L + S as the total angular momentum of the particle, $J^2$ and $H_D$, as well as the parity operator P. Written out as an equation, this results in:

$$\psi_{E\kappa m_j} = \frac{1}{r}\begin{pmatrix} P_{E\kappa}(r)\, Y^{j,m_j}_{l\frac{1}{2}} \\ iQ_{E\kappa}(r)\, Y^{j,m_j}_{l'\frac{1}{2}} \end{pmatrix} \quad mit \quad Y^{j,m_j}_{l\frac{1}{2}} = \sum_{m_l, m_s} C\left(l\,\frac{1}{2}\,j;\, m_l m_s m\right) Y_{l,m}(\theta, \varphi) \chi_{\frac{1}{2}, m_s}$$

The spin angle functions $Y^{j,m_j}_{l\frac{1}{2}}$ are the products of the spherical harmonic functions and the spin functions $\chi_{\frac{1}{2}, m_s}$. The evolution coefficients C are known from the literature as Clebsch-Gordan coefficients. The radial functions $P_{E\kappa}(r)$ and $Q_{E\kappa}(r)$ for particles and antiparticles result in

$$\left[\frac{d}{dr} - \frac{\kappa}{r}\right] P_{E\kappa}(r) = \frac{E + m_0 c^2 - V(r)}{\hbar c} Q_{E\kappa}(r) \tag{15}$$

$$\left[\frac{d}{dr} + \frac{\kappa}{r}\right] Q_{E\kappa}(r) = \frac{E - m_0 c^2 - V(r)}{\hbar c} P_{E\kappa}(r) \tag{16}$$

Hence $\kappa$ is defined as: $\quad \kappa = \pm\left(j + \frac{1}{2}\right) = \begin{cases} l & for\ j = l - \frac{1}{2} \\ -(l+1) & for\ j = l + \frac{1}{2} \end{cases}$

One can imagine that the Dirac equation, in its complexity and variety of being able to represent various particle states including their antiparticles, can be a complex task to parameterize. For electrons in the K-shell with n=1, l=0, J=0, $\frac{1}{2}$, however, the equation can be simplified in a pleasing way. Thus, with the requirement for asymptotic behavior for $\to \infty$ and with the substitutions

$$P_{E\kappa}(r) = p_E(r)e^{-a_0 r} \quad and$$

$$Q_{E\kappa}(r) = q_E(r)e^{-a_0 r}$$

with $a_0 \in \mathbb{R}^+$ one obtains convergent solutions that resemble the wave function of the 1s orbital from the Schrödinger equation. The energy levels then result in:

$$E^D_{n,j} = m_0 c^2 \left[1 + \left(\frac{Z\alpha}{n - j - \frac{1}{2} + \sqrt{\left(j + \frac{1}{2}\right)^2 - (Z\alpha)^2}}\right)^2\right]^{-\frac{1}{2}} \tag{17}$$

Where $\alpha = \frac{e^2}{4\pi\varepsilon_0 \hbar c}$ is the fine structure constant. If n=1 and j=$\frac{1}{2}$, the equation simplifies significantly to:

$$E^D_{n,j} = m_0 c^2 \sqrt{1 - (Z\alpha)^2} \tag{18}$$

For hydrogen. For elements with Z>1 and at least two electrons with 1s$^2$ configuration in the lower-energy singlet state we obtain [27]:

$$E_{n,j}^D = m_0 c^2 \left[ 1 + \left( \frac{Z\alpha}{\frac{1}{2} + \sqrt{\frac{1}{4} - (Z\alpha)^2}} \right)^2 \right]^{-\frac{1}{2}}$$

Now the Coulomb-Casimir potential can be extended with the new expression from equation (13):

$$V(r) = V_{Coulomb} + V'_{Casimir} = -\frac{Ze^2}{4\pi\varepsilon_0} \frac{1}{|r|} - \frac{\hbar c \pi^3 r_k{}^2}{240(\sqrt{2}\varepsilon\alpha)^3} \left( e^{-\frac{3}{2}} + \frac{1}{4} \right) \left[ Erf\left( \frac{r - r_k}{\sqrt{2}\varepsilon\alpha} \right) - 1 \right]$$

The Dirac equation is solved numerically. This was carried out using Matlab with the ODE45 solver [51] and the code is documented in the annex. It results in two wave functions $p_E(r)$ and $q_E(r)$, which are typically represented as an array with $10^6$ coefficients for further processing. The high resolution is required to resolve both the nucleus area with $<10^{-15}$m together with the outer area of the orbital with $10^{-11}$m in sufficient accuracy. This is necessary to avoid numerical interferences and artifacts.

## 5. Numerical solutions of the Dirac equation with the Casimir-Coulomb potential

In order to examine the assumption more closely as to whether the Casimir effect actually influences the inner electrons, the electrons of the K shell were specifically analyzed as described in the previous chapter, compared with the literature [29,30,31,32,33,34,35,36,37,38,39,40] and the result shown in the following figures. For better comparability, the Dirac equation was solved exemplarily for $H_1^1$, $Cu_{29}^{65}$ and $Pb_{82}^{206}$ first only with Coulomb-potential and the second time with the combination of Coulomb and Casimir potential, i.e. quasi-classically and with the presented model respectively. In addition, the quadrupole moments for the respective elements from [41,42,43,44,45,46,47,48,49,50] were considered in the calculation in order to investigate their effect on the energy.

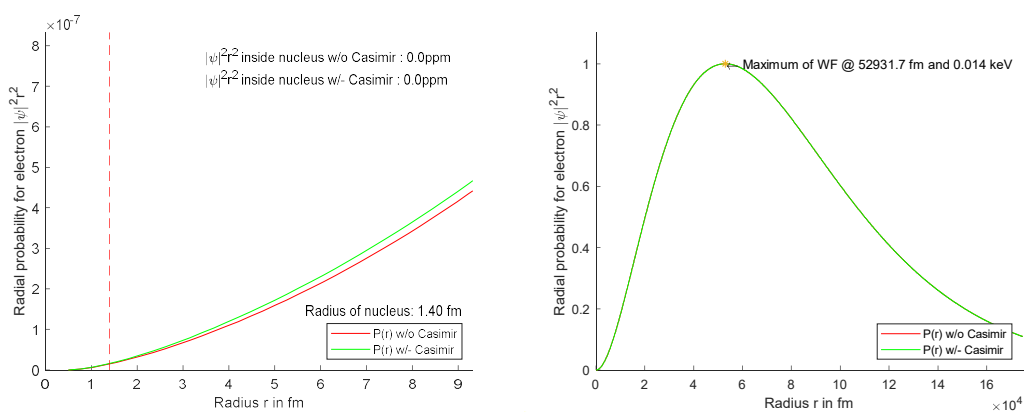

Fig.4: Radial probability $|\psi|^2 r^2 dr$ of the K-shell electron of $H_1^1$ plotted against the nuclear distance in fm; the near field of the nucleus is shown on the left, the overall view on the right. The red line represents the Dirac solution without Casimir effect, the green line the result with Casimir effect. As can be easily seen, the Casimir effect has virtually no influence on the system. The nucleus radius was assumed to be 1.4 fm and the binding energy is -13.6eV. Source: Calculation by Matlab script.

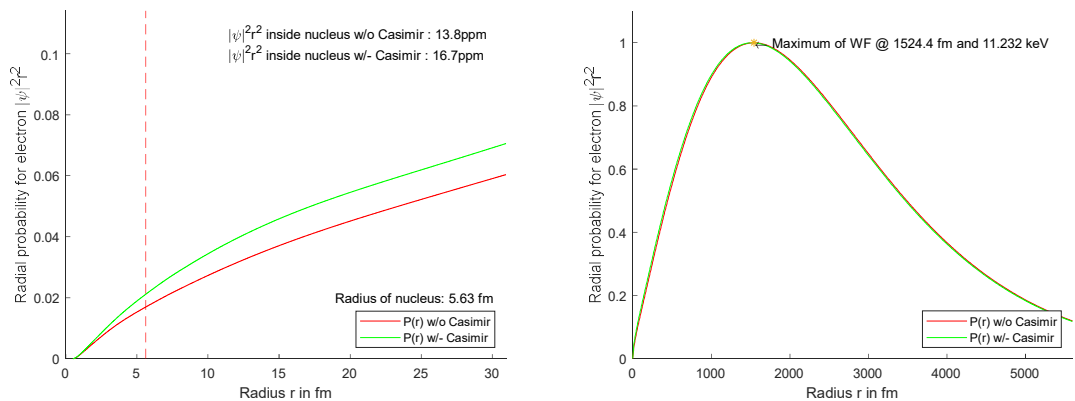

Fig.5: Radial probability $|\psi|^2 r^2 dr$ of the K-shell electron of $Cu_{29}^{65}$ plotted against the nuclear distance in fm; the near field of the nucleus is shown on the left, the overall view is found on the right side. The red line represents the Dirac solution without Casimir effect, green shows the influence of the Casimir effect. A slight shift of the probability towards the nucleus can be seen. The nucleus radius was assumed to be 5.63 fm and the binding energy is calculated as 11.2 keV. Source: Calculation by Matlab script.

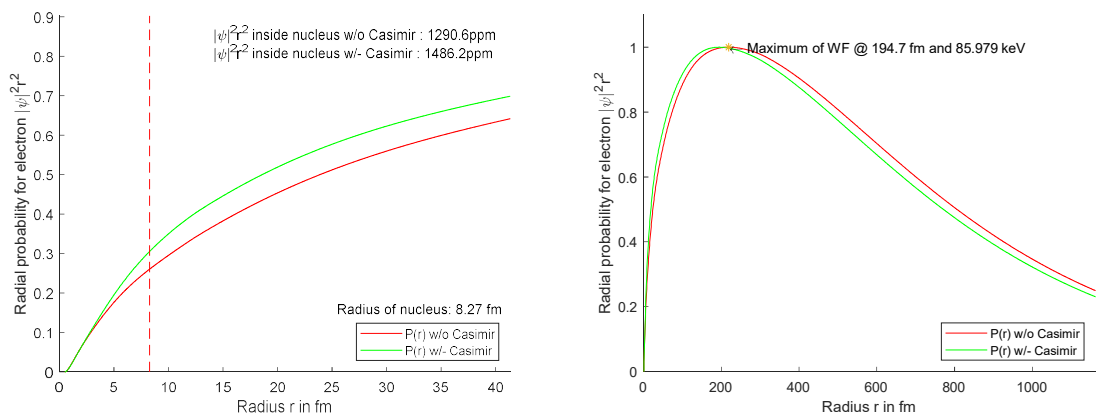

Fig.6: Radial probability $|\psi|^2 r^2 dr$ of the K-shell electron of $Pb_{82}^{206}$ plotted against the nuclear distance in fm; on the left the near-field of the nucleus is shown, the right side gives an overall view. The red line represents the Dirac solution without Casimir effect, green shows the influence of the Casimir effect. A clear shift of the probability towards the nucleus can be seen. The nucleus radius was assumed to be 8.27 fm and the binding energy is calculated as 86 keV. Source: Calculation by Matlab script

One recognizes an increasingly significant influence of the Casimir effect on the K-shell electron with increasing atomic number Z. For hydrogen and copper there are no, or only negligible deviations in the electron probability at the nucleus of 0 or 16.7 instead of 13.8 ppm – a factor of 1.2, but negligible in absolute terms - in the case of Pb there is an increase from 1291 ppm to 1486 ppm by a factor of 1.15. This may seem small, but it clearly shows the influence of the modified potential on the orbitals. Even more remarkable are the effects on the binding energy.

### 6. determination of the energy by first order perturbation calculation

The binding energy $E_K + E_C^1$ was determined using the well-known first-order perturbation approach, i.e. the numerically calculated electron wave functions were multiplied with the original Casimir potential using vector multiplication and the result was evaluated. The following therefore applies:

$$E_C^1 = \langle \psi^0 | V_{Casimir} | \psi^0 \rangle \qquad (21)$$

The Casimir potential from equation (27) was used. The result is shown in Table 1 for the three example elements listed above and compared with the literature [29-37,39,40]. It is important to note

| element | Quadrupole mom. $Q_b$ | $E_K$ (Dirac) | $E_C^1$ | $E_K + E_C^1$ | $E_K$/ literature | Δ vs. $E_K$ Dirac | Δ vs. $E_K + E_C^1$ |
|---|---|---|---|---|---|---|---|
| $H_1^1$ | 0,003 | 13,6 eV | <1*10⁻²³ eV | 13,6 eV | 13,6 eV | 0% | 0% |
| $Cu_{29}^{65}$ | 0,22 | 11,57 keV | 0,81 keV | 11,5 keV | 8,98 keV | 29% | 28% |
| $Pb_{82}^{206}$ | 0,23 | 101,58 keV | 12,9 keV | 88,7 keV | 88,01 keV | 15% | **1%** |

Table 1: Comparison of the binding energy and the first-order correction terms as well as the relative deviations for the elements hydrogen, copper and lead selected as examples

here that the values of the Dirac equation refer primarily to one-electron systems, i.e. the elements are practically in the highly ionized state down to the K-electron shell. The literature values, on the other hand, have been measured on the neutral atom, so the Coulomb and exchange interactions between the multiple electrons have to be taken into account, and this explains the higher energies, especially for the light elements. However, both Coulomb and exchange interactions lose influence in relation to the atomic number for the heavy elements [19,20] so that this alone cannot explain the energy difference. Here, the Casimir effect fits into the picture like a missing piece of the puzzle.

One can see that the energy correction for hydrogen is effectively zero, as the Casimir potential is virtually ineffective due to the distance of the electron from the nucleus. In the case of copper with an average radius of approx. 1.5pm, there is a small effect, which, however, cannot fully explain the discrepancy between the measured binding energy and the $E_K$ resulting from the Dirac equation.

Finally, in the case of lead, the model with Casimir effect can close the gap almost completely; with an average radius of approx. 200 fm (see Figure 6), the probability of the electron is so close to the nucleus that the Casimir forces described become fully effective. The Casimir effect thus acts in the form of a potential well, which slightly reduces the binding energy of the Coulomb potential, i.e. slightly expands the potential well, but at the same time shifts the probability of the electron's position towards the nucleus due to its shape, provided the electron is close enough to the nucleus. This effect will therefore be particularly effective for heavy elements. One may ask why the attractive effect can simultaneously lower the binding energy. Now the Casimir force acts near the edge of the nucleus, especially for small $r > r_k$; within the nucleus, however, $F_{Casimir}$ gives no contribution, i.e. according to the structure of a potential well, but with a broadened wall. For the electron, this effect acts like a broadening of the Coulomb potential near the nucleus, as can be clearly seen in Figure 3. Consequently, the effective potential does not become lower, but wider, which leads to the seemingly paradoxical situation that the binding energy is reduced even though the electron is closer to the nucleus.

### 7. The influence of the Casimir effect on the binding energy of isotopes

The situation is particularly interesting when considering different isotopes of a heavy element. In Table 2, six different mercury isotopes with their respective quadrupole moments were calculated with respect to K-shell binding energy [29-37,39,40,43], evaluated and compared with the literature. It should be noted that some of the quadrupole moments were generated by selective preparation of the isotope and may have had a lifetime in the ps range - thus, of course, they do not represent their natural occurrence. In addition, the nucleus was estimated at 8.2 fm radius and presumably also only partially corresponds to the real radius of the different isotopes. Data for this are also only available to a very limited extent in the literature. Consequently, the comparison of the calculated binding energies with the mean distribution of the literature value is only possible to a limited extent and a systematic deviation can be expected. Nevertheless, the calculation clearly shows how sensitive the binding energy reacts to nuclear radius and shape, quadrupole moment and, last but not least, atomic mass. An isotope distinction based solely on the measurement of the $K_\alpha$ line thus seems obvious, provided the spectrometric resolution is sufficient; presumably this must be below 10nm@ 80keV for the mercury $K_\alpha$ line. However, this still appears to be quite demanding concerning the technological requirements of the spectrometer.

| element | Quadrupole mom. $Q_b$ [barn] | $E_K + E_C^1$ | $E_B$/literature | $\Delta E$ vs. $E_K + E_C^1$ [eV] | $\Delta E$ vs. $E_K + E_C^1$ [%] |
|---|---|---|---|---|---|
| $Hg_{80}^{198}$ | +0,68 | 83,18 keV | 83,102 keV | 78 eV | 0,09 |
| $Hg_{80}^{199}$ | +1,2 | 83,93 keV | 83,102 keV | 828 eV | 0,99 |
| $Hg_{80}^{200}$ | +0,96 | 83,50 keV | 83,102 keV | 398 eV | 0,48 |
| $Hg_{80}^{201}$ | +0,387 | 82,48 keV | 83,102 keV | -622 eV | -0,75 |
| $Hg_{80}^{202}$ | +0,87 | 83,23 keV | 83,102 keV | 128 eV | 0,15 |
| $Hg_{80}^{204}$ | +0,4 | 82,30 keV | 83,102 keV | -802 eV | 0,96 |

Table 2: Comparison of the calculated binding energies for six mercury isotopes and their quadrupole moments (from [43]). The absolute and relative deviation of the binding energies for each isotope from the literature value is shown. The systematic deviation from the literature value in relation to the abundance of the isotopes is presumably due to inaccuracies of the nuclear radius in the calculation

### 8. Evaluation of the results and consequences for the binding energy

It is noteworthy that this counterintuitive effect is decisively influenced by the structure of the nuclear shell. As can easily be seen from equations (13) and (27), the Casimir force will increase the more the boundary zone approaches a spherical shell. In the ideal case of an infinitely sharp boundary surface, (13) would return to (5) at the boundary transition.

Conversely, this also means that the multipole moments of the nucleus influence the binding energy of the K-shell electron, and that the binding energy - isotope-specifically - can report details about the nuclear structure. Isotope-specific because a variation of the neutron number cannot influence the charge number Z, but it can influence the nuclear symmetry and thus the multipole moments [46]. If this can be confirmed by experiment, analysis methods using the investigation of the $K_\alpha$ —line would open up a very easy and direct path to structural analysis of the heavy atomic nuclei. In this case, it would be interesting to analyze this relationship in detail.

For the sake of simplicity, only the quadrupole moment of the elements was considered here; for a complete analysis, all multipole moments, or at least all those with a significant contribution, should be considered. In addition, it should be checked experimentally whether- and how strong the binding energy of the K-shell varies isotope-specifically and thus deviates from the predictions of the classical atomic models.

Finally, binding energies for all elements from $H_1^1$ to $No_{102}^{248}$ were derived using this method to obtain an overall picture of the phenomenon. The result can be seen in Figure 7. Here, the percentage deviation of the measured energy value from the literature is compared with the calculated binding energy from the Dirac equation $E_K$ without or with energy perturbation term $E_C^1$ and plotted as the difference. The elements Te, Bi, Cm, Bk, Cf, Fm, Md and No could not be considered because either value for the binding energy or the quadrupole moment were not available. While for $E_K$ there is consistently a shift towards smaller values than known from the literature value due to reasons discussed before, the inclusion of the perturbation term $E_C^1$ shows a significantly smaller deviation from the (green) zero line, especially for the heavy elements, and thus a better agreement with the values from the experiments. For some elements such as lead, Polonium and Radon, the calculated values agree very well with the literature; for the other values, it remains to be clarified whether the deviations can be reduced by adding further multipole elements such as octupole or higher multipole moments or more complex deviations from the spherical symmetry, or whether other effects that have not yet been considered are effective.

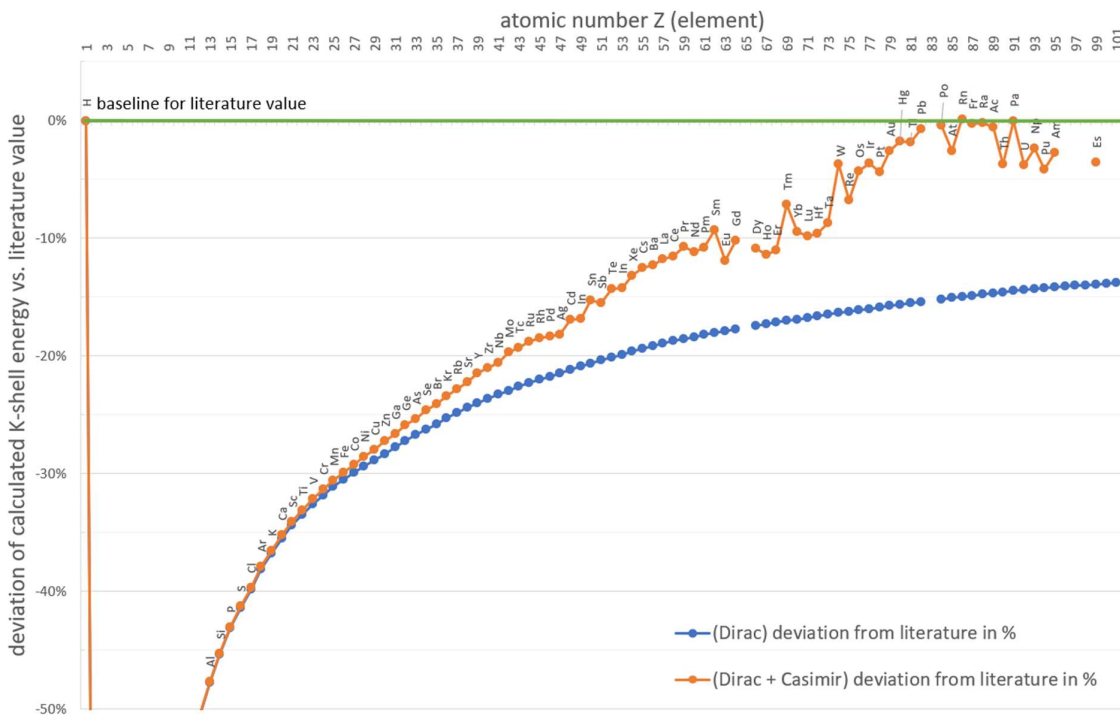

Fig. 7: Influence of the Casimir potential on the binding energies of the K-shell electron as a percentage deviation from the measured literature value (green zero line). The values from the Dirac equation are plotted in blue, orange values contain the influence of the Casimir potential. It can be clearly seen that the Casimir effect only becomes effective in the heavier elements, where it is clearly influenced by the multipole moments, here in particular the quadrupole moment

It should not be concealed that this work offers yet another aspect: Provided that the residence probabilities and binding energies can be better explained by the presented effect, there is also the possibility to determine the decay constants and transition probabilities of electron capture more precisely. Since $|\psi_e(r)|^2 r^2 dr$ shifts significantly in some cases due to the Casimir effect, this must of course be included in the equations for the transition probability in addition to the nuclear matrix element and the energy conditions. Since a higher probability is consistently observed near the nucleus, it is to be expected that the probability of electron capture - if energetically permitted - also increases. This raises the question of the lifetime of the (super)heavy elements, the so-called islands of stability, which are assumed to exist in the range beyond Z=150. If the Casimir effect is effective near the nucleus, the existence of these islands becomes at least less probable.

### 9.  Summary and Outlook

In the present work, the Casimir effect in the near-nucleus environment and its impact on the elec­trons of the K-shell was considered in the form of a numerical model for all elements, especially for heavy elements. An approach for the Casimir potential for non-ideal surfaces was derived. In particu­lar, nucleus surfaces with non-spherical components, in this case in first approximation with quadru­pole components, were considered. Solving the Dirac equation with this combined potential shows that the solutions for the wave functions of the K-shell electrons have a significantly higher residence probability at the nucleus than with the pure Coulomb potential. In addition, it could be shown by first-order perturbation calculations that with this approach the calculated binding energies, especially for heavy elements, almost agree with the measured literature values from spectroscopy. Probably the most important finding is that - if the model presented here is correct - the nature of the nuclear surface, in particular the deviation from the spherical nuclear geometry, has a much stronger influence on the electron binding energies than previously assumed. This may open up the possibility of using spectroscopy of the binding energies to gain easier access to the investigation of the nucleus structure and the nature of its surface. Last but not least, if confirmed, the model will allow greater accuracy in calculating the transition probabilities and the decay constants of electron capture. The shifted prob­abilities of the electron within the nuclear structure, together with the nuclear matrix element and the energy conditions, have a direct effect on the corresponding half lifetime of the isotope under consideration. Furthermore, it seems worthwhile to check whether the literature values of other ele­ments or isotopes can also be reproduced by adding further multipole elements. And finally, it should be analyzed in detail whether - and how far - $E_K$ has an isotope-specific component.

**10. Annex: Integral calculation using the convolution theorem**

The integration of equation 10 cannot be solved straight forward, but there is a way via the convolution theorem. This states that in a Fourier transformation, for example, a convolution in the time domain corresponds to a multiplication in the frequency domain. The desired function is obtained by reverse transformation. The following therefore applies:

$$\mathcal{F}[(f * F)(r)](k) = f(k) \cdot F(k) = h(k) \quad \Leftrightarrow \quad \mathcal{F}^{-1}[h(k)](r) = h(r) \tag{22}$$

For simplification and to avoid phase factors, we substitute $r' = r - r_k$. In addition, $f(r) \rightarrow \delta(r)$ must apply for the transition $\varepsilon\alpha \rightarrow 0$ In particular, the following then applies:

$$\lim_{\varepsilon\alpha \rightarrow 0} \int f(\zeta - r')F(\zeta)d\zeta = -\frac{\hbar c \pi^3 r_k{}^2}{60} \int \frac{\delta(\zeta - r')}{(\zeta)^4}d\zeta = \langle F \rangle A \tag{23}$$

To ensure the identity in (20), we add the parameter $\lambda =$const to the distribution function f. In addition, we substitute $\chi := \sqrt{2}\varepsilon\alpha$ for simplification, where $\chi$ is constant but freely selectable. The two functions from (6) and (7) act as a starting point

$$F(r) = -\frac{\hbar c \pi^3 r_k{}^2}{60 r'^4}$$

and

$$f(r) = \lambda e^{-\frac{r'^2}{\chi^2}} \ with \int_{\mathbb{R}} f(r)dr = 1$$

The Fourier transformed functions can be expressed respectively:

$$F(k) = -\frac{\hbar c \pi^3 r_k{}^2}{60} \int \frac{e^{-ikr'}dr'}{r'^4} = -\frac{\hbar c \pi^3 r_k{}^2}{360}\sqrt{\frac{\pi}{2}}k^3 sgn(k)$$

and

$$f(k) = \lambda \int e^{-\frac{r'^2}{\chi^2}}e^{-ikr'}dr' = \lambda\frac{\chi}{\sqrt{2}}e^{-\frac{1}{4}\chi^2 k^2}$$

The following then applies to the reverse transformation:

$$\mathcal{F}^{-1}[h(k)](r') = -\lambda\frac{\hbar c \pi^3\sqrt{\pi}r_k{}^2}{720}\chi \int k^3 sgn(k)e^{-\frac{1}{4}\chi^2 k^2}e^{ikr'}dk = h(r') \tag{24}$$

$$\Leftrightarrow h(r') = -\lambda \frac{\hbar c \pi^3 \sqrt{\pi} r_k{}^2}{720} \chi \cdot Re\left\{\left|-\frac{8\sqrt{2}}{\chi^7} e^{-\frac{r'^2}{\chi^2}} \left[\frac{3}{2}\chi^2 r' - r'^3\right]\right|\right\} \qquad (25)$$

As mentioned above, at the $\chi \to 0$ limit transition $h(r')|_{r' \to 0}$ must return to the original expression for the Casimir force. Let us therefore consider:

$$\lim_{\chi \to 0}\left[-\lambda \frac{\hbar c \pi^3 \sqrt{\pi} r_k{}^2}{720} \chi \cdot \left|-\frac{8\sqrt{2}}{\chi^7} e^{-\frac{r'^2}{\chi^2}} \left[\frac{3}{2}\chi^2 r' - r'^3\right]\right|\right] \overset{\text{def}}{=} -\frac{\hbar c \pi^2}{240 r'^4} A\Bigg|_{r' \to \chi}$$

With $e^{-\frac{r'^2}{\chi^2}} \to \delta(r')$ and $\lim_{\chi \to 0} \frac{3}{2}\chi^2 r' \to 0$ the equation can be simplified to

$$\Leftrightarrow \lim_{\chi \to 0}\left[-\frac{\hbar c \pi^3 r_k{}^2}{240}\lambda \frac{8\sqrt{2\pi}}{3}\int \frac{\delta(r') r'^3}{\chi^6} dr'\right]\Bigg|_{r' \to \chi} = \lim_{\chi \to 0}\left[-\frac{\hbar c \pi^3 r_k{}^2}{240}\lambda \frac{8\sqrt{2\pi}}{3}\frac{1}{\chi^3}\right]\Bigg|_{r' \to \chi}$$

This results in the mathematical limit value

$$\Leftrightarrow \lim_{\chi \to 0}\left[-\frac{\hbar c \pi^3 r_k{}^2}{240}\frac{1}{\chi^4}\lambda \frac{8\sqrt{2\pi}}{3}\chi\right]\Bigg|_{r' \to \chi} = -\frac{\hbar c \pi^2}{240 r'^4} A\Bigg|_{r' \to \chi} \quad f\ddot{u}r\ \lambda = \frac{3}{8\sqrt{2\pi}\chi}$$

Back substitution with $r' = r - r_k$ and $\chi := \sqrt{2}\varepsilon\alpha$ now finally results in the analytically derived term for the Casimir force at the diffuse interface of a spherical shell:

$$\Leftrightarrow h(r) = F_{Casimir}(r) = -\frac{\hbar c \pi^3 r_k{}^2}{240(\sqrt{2}\varepsilon\alpha)^7} e^{-\frac{(r-r_k)^2}{2(\varepsilon\alpha)^2}} |3(\varepsilon\alpha)^2(r-r_k) - (r-r_k)^3| \qquad (26)$$

To obtain $V_{Casimir}$, the integration over r must be carried out piecewise due to the sgn function:

$$V_{Casimir}(r) = -\frac{\hbar c \pi^3 r_k{}^2}{240(\sqrt{2}\varepsilon\alpha)^5} \cdot \begin{cases} -e^{-\frac{(r-r_k)^2}{2(\varepsilon\alpha)^2}}\left(\frac{1}{2}(r-r_k)^2 - \frac{1}{2}(\varepsilon\alpha)^2\right) & f\ddot{u}r\ r > r_k + \sqrt{3}(\varepsilon\alpha) \\ e^{-\frac{(r-r_k)^2}{2(\varepsilon\alpha)^2}}\left(\frac{1}{2}(r-r_k)^2 - \frac{1}{2}(\varepsilon\alpha)^2\right) + (\varepsilon\alpha)^2 e^{-\frac{3}{2}} & f\ddot{u}r\ r_k < r < r_k + \sqrt{3}(\varepsilon\alpha) \\ -e^{-\frac{(r-r_k)^2}{2(\varepsilon\alpha)^2}}\left(\frac{1}{2}(r-r_k)^2 - \frac{1}{2}(\varepsilon\alpha)^2\right) + (\varepsilon\alpha)^2\left(e^{-\frac{3}{2}} + \frac{1}{2}\right) & f\ddot{u}r\ r_k - \sqrt{3}(\varepsilon\alpha) < r < r_k \\ e^{-\frac{(r-r_k)^2}{2(\varepsilon\alpha)^2}}\left(\frac{1}{2}(r-r_k)^2 - \frac{1}{2}(\varepsilon\alpha)^2\right) + 2(\varepsilon\alpha)^2\left(e^{-\frac{3}{2}} + \frac{1}{4}\right) & f\ddot{u}r\ r < r_k - \sqrt{3}(\varepsilon\alpha) \end{cases} \qquad (27)$$

It is important to ensure that the function remains continuous and differentiable at the connection points. This results in the integration constants listed in (25). The depth of the potential well is given by

$$\Delta V_{Casimir} = V_{Casimir}(\infty) - V_{Casimir}(0) = -\frac{\hbar c \pi^3 r_k{}^2}{240(\sqrt{2}\varepsilon\alpha)^3}\left(e^{-\frac{3}{2}} + \frac{1}{4}\right) \qquad (28)$$

It was taken into account that converges to zero for infinity: $V_{Casimir}(\infty) \to 0$ and that

$$V_{Casimir}(r = 0) = \left[ -\frac{\hbar c \pi^3 r_k{}^2}{240(\sqrt{2}\varepsilon\alpha)^5} e^{-\frac{(r - r_k)^2}{2(\varepsilon\alpha)^2}} \left( \frac{1}{2}(r - r_k)^2 - \frac{1}{2}(\varepsilon\alpha)^2 \right) + 2(\varepsilon\alpha)^2 \left( e^{-\frac{3}{2}} + \frac{1}{4} \right) \right]\Bigg|_{r=0}$$

$\Rightarrow V_{Casimir}(r = 0) \rightarrow -\frac{\hbar c \pi^3 r_k{}^2}{240(\sqrt{2}\varepsilon\alpha)^3} \left( e^{-\frac{3}{2}} + \frac{1}{4} \right)$ remains valid.

$V_{Casimir}$ can be approximated very well by the $Erf$ function, whereby the potential depth is scaled as described above. This results in $V'_{Casimir}$ as already mentioned in equations (12) and (18):

$$V'_{Casimir}(r) = -\frac{\hbar c \pi^3 r_k{}^2}{240(\sqrt{2}\varepsilon\alpha)^3} \left( e^{-\frac{3}{2}} + \frac{1}{4} \right) \left[ Erf\left( \frac{r - r_k}{\sqrt{2}\varepsilon\alpha} \right) - 1 \right] \qquad (29)$$

### 11. Matlab-Code for the numerical calculations

The Matlab script used essentially consists of three parts: 1. the definition of the constants, 2. the definition and solution of the coupled differential equations using the ODE45 solver and 3. the evaluation of the wave functions and the determination of the energy by first-order perturbation calculation. The commentary is largely self-explanatory; it is important to ensure that the numerical resolution of the wave function is chosen fine enough to avoid numerical artifacts. Care should also be taken to ensure that the selected potential is mathematically smooth, insofar as it can be represented numerically at all. The results can be further processed, displayed graphically or saved as desired.

```matlab
%% Numerische Simulation der atomaren K-Schale mit Hilfe der Dirac-Gleichung
%% Es wird die Aufenthaltswahrscheinlichkeit eines Elektrons auf der K-Schale für beliebige Elemente simuliert
clear all; close all; clc
tic;
% Definiere die Konstanten
    hquer           = 6.62607015e-34/(2*pi);                        % Js
    C               = 299792458;                                   % m/s
    e_e             = 1.602176634e-19;                             % Elementarladung in Coulomb
    m_e             = 9.1093837015e-31;                            % kg -> Elektronenmasse
    m_P             = 1.67262192369e-27;                           % kg -> Protonenmasse
    r_P             = 0.8409e-15;                                  % Protonenradius in m
    m_N             = 1.67492749804e-27;                           % kg -> Neutronenmasse
    Epsilon0        = 8.8541878128e-12;                            % A*s/(V*m)
    Atommasse       = 197;                                         %
    Z_Ladung        = 79;                                          % Ladungszahl des Kerns, hier Pb203
    QP_Moment       = +0.547;                                      % Qp
    m_e_red         = m_e/(1+ m_e/(Z_Ladung*m_P+(Atommasse-Z_Ladung)*m_N));  % Berechne die reduzierte Masse des Elektrons
    K_1             = Z_Ladung * e_e^2 / (4 * pi * Epsilon0);      % Zur Vereinfachung
    K_2             = hquer^2/(2*m_e_red);                         % Ebenfalls zur Erleichterung
    K_alpha         = e_e^2 / (4 * pi * Epsilon0 * hquer * C);     % Sommerfeldsche Feinstrukturkonstante
    E_0             = m_e*C.^2;                                    % Ruheenergie des Elektrons
    E_Proton        = 8.0e6*e_e;                                   % Bindungsenergie des Protons
    E_H             = e_e^4*m_e/(8*Epsilon0^2*(hquer^2*pi)^2);     % Bindungsenergie für Wasserstoff
    Wandbereich     = hquer/sqrt(2*m_P*E_Proton);                 % Wandbereich des Potentialtopfes
    r_Kern          = 7.69e-15;%1.21e-15 * Atommasse^(1/3);       % Kernradius
    a_Kern          = 0.5e-15;                                     % m: Randunschärfe des Kerns
% Konstanten für das Casimir-Potential
% Kern-Quadrupolmoment in barn
    Delta_R_R       = 5*QP_Moment/(4*Z_Ladung*r_Kern)*1e-28;      % Delta = DeltaR/R liegt typischerweise zw. 0.01 und 0.1
    a_Kern_Casimir  = sqrt(2)*(Delta_R_R + Wandbereich*2);        % m: Für die Kugelhomogenität wird das Kern-Quadrupolmoment herangezogen
% Bestimmung der Energie mit Z, n=1, l=0, s = 1/2 Laplace und Bohrsche Radien für 1s, 2s und 3s Orbitale
    Energy_1s       = ((1 - (Z_Ladung .* K_alpha).^2)).^(0.5) .* E_0;        % Energie aus der Dirac-Gleichung
% Energie aus der Dirac-Gleichung für alle anderen Elemente
    end
    beta_a_0_1s     = 2 * K_2 / K_1;                              % 1. Bohrscher Radius
% Definiere die Numerischen Parameter
    r_start         = a_Kern;                                     % Startpunkt - kann beim Wood Saxon Potential 0 sein
    spacing         = 1000000;                                    % Anzahl der zu berechnenden Punkte für r
    r_end           = beta_a_0_1s * 50;                           % Endpunkt außerhalb der K-Schale
    r_Span          = linspace (r_start, r_end, spacing);        % Definiere den Vektor r
    delta_r         = (r_end - r_start) / spacing;

% Skalierung
    graph_scale     = 1e15;
% Anfangswerte für die DGL
    psi0 = [0;1]; % Initial values of psi
% Definiere die DGL
kappa = 0; % Kappa kann 0 oder -1 sein
r_Kern_index = max(find(r_Span<=r_Kern));

% Allokiere den Speicherplatz
P_psi       = zeros(spacing,2); P_psi_2    = zeros(spacing,2);
r_max       = [0,0]; A_P_psi_1s = [0,0]; P_psi_Kern = [0,0];
% Berechne den Normierungsfaktor
% Definiere die Casimir-Kraft
    z           = exp(-(r_Span-r_Kern).^2./(a_Kern_Casimir.^2)); norm_B = 1./(delta_r * trapz(z));
F_Casimir = -norm_B*hquer*C*pi^3.*r_Kern.^2/(240.*a_Kern_Casimir.^7).*z.*abs(1.5.*a_Kern_Casimir.^2.*(r_Span-r_Kern)-(r_Span-r_Kern).^3);
    V_Casimir = -cumtrapz(F_Casimir); V_Casimir = V_Casimir - max(V_Casimir);
K_Casimir = (hquer*C*pi^3*r_Kern^2/(240*a_Kern_Casimir^3)*(exp(-1.5).*.25));
V_Casimir = V_Casimir.*abs(min(V_Casimir)).*K_Casimir; K_erf      = K_Casimir/2;
% Rechne dann einmal mit und einmal ohne das Casimir-Potential
for i=1:2
    if i==1
        casimir_schalter = 0;                                    % Zuerst ohne Casimir Potential rechnen
    else
        casimir_schalter = 1;                                    % Dann mit Casimir Potential rechnen
    end
    %odefun = @(r,psi) [ ((Energy_1s + m_e.*C.^2 - (V)) .* (-beta_a_0_1s./hquer./C)) .* psi(2) + kappa.*psi(1)./r ;
    %                    ((Energy_1s - m_e.*C.^2 - (V)) .* (-beta_a_0_1s./hquer./C)) .* psi(1) - kappa.*psi(2)./r];
    odefun = @(r,psi) [ ((Energy_1s + m_e.*C.^2 - (- K_1./r + casimir_schalter .* K_erf.*(-1 + erf((r - r_Kern)./a_Kern_Casimir) )) ) .* (-beta_a_0_1s./hquer./C)) .* psi(2) + kappa.*psi(1)./r ;
```

```matlab
                            ((Energy_1s - m_e.*C.^2 - (- K_1./r + casimir_schalter .* K_erf.*(-1 + erf((r - r_Kern)./a_Kern_Casimir) )) .* ...
    (-beta_a_0_1s./hquer./C)) .* psi(1) - kappa.*psi(2)./r];
        [r_result, psi_result] = ode45(odefun, r_Span, psi0);
        % Normiere die Wellenfunktion und berechne die Aufenthaltswahrscheinlichkeit des Elektrons
        P_psi(:,i)    = psi_result(:,1).*exp(-r_Span'./beta_a_0_1s);       % Regularisierung für r --> oo
        A_P_psi_1s(i) = 1/sqrt(abs(P_psi(:,i)'*P_psi(:,i)));              % Normierungsfaktor
        P_psi_2(:,i)  = abs(P_psi(:,i).^2 .* A_P_psi_1s(i)^2);           % Radiale Aufenthaltswahrscheinlichkeit
        % Aufenthaltswahrscheinlichkeit des Elektrons im Kern
        P_psi_Kern(i) = trapz(P_psi_2(1:r_Kern_index,i));                % Bestimme die Aufenthaltswahrscheinlichkeit des Elektrons im
Kern
        % Finde das Maximum der Wellenfunktion
        P_index = find(P_psi_2(:,i)==max(P_psi_2(:,i)));                 % Bestimme den Index für das Array
        r_max(i)  = [ r_result(P_index)];
end
% Rechne die Energiedifferenz aus gemäß <psi_0|E_casimir|psi_0>
E_SR1 = abs(P_psi(:,1)'*(V_Casimir'.*P_psi(:,1)) * A_P_psi_1s(1)^2);
fprintf('Bindungsenergie der K-Schale aus der Dirac-Gleichung %3.2f  keV \n',(Energy_1s-E_0+5/8*Z_Ladung*E_H)/e_e/1000);
fprintf('E_Dirac + E_Störung %3.2f  keV \n',(Energy_1s-E_0+E_SR1+5/8*Z_Ladung*E_H)/e_e/1000);
toc; % Beende CPU-Zeitmessung;
```